\documentclass[twocolumn,aps,prl,superscriptaddress,amsmath,amssymb,letterpaper]{revtex4}

\makeatletter
\renewcommand{\p@subsection}{}
\renewcommand{\p@subsubsection}{}
\makeatother
\usepackage{graphicx}
\usepackage{dcolumn}
\usepackage{bm}
\usepackage{xcolor}
\usepackage{soul}  
\usepackage{enumitem} 

\newcommand*{\MIT }{Massachusetts Institute of Technology, Cambridge, Massachusetts 02139, USA}
\newcommand*{\ODU}{Old Dominion University, Norfolk, Virginia 23529, USA}
\newcommand*{\JLAB}{Thomas Jefferson National Accelerator Facility, Newport News, Virginia 23606, USA}
\newcommand*{\TAU }{School of Physics and Astronomy, Tel Aviv University, Tel Aviv 69978, Israel}
\newcommand*{\Penn}{Pennsylvania State University, University Park, PA, 16802, USA}
\newcommand*{\GW}{George Washington University, Washington, D.C., 20052, USA}
\newcommand*{\HUJI}{The Racah Institute of Physics, The Hebrew University, Jerusalem 9190401, Israel}
\newcommand*{\NRCN}{Nuclear Research Center Negev, Be'er Sheva 84190, Israel}

\begin{document}


\title{Generalized Contact Formalism Analysis of the $^4$He$(e,e'pN)$ Reaction}

\author{J.R. Pybus}
\affiliation{\MIT}
\author{I. Korover}
\affiliation{\NRCN}
\affiliation{\TAU}
\author{R. Weiss}
\affiliation{\HUJI}
\author{A. Schmidt}
\affiliation{\GW}
\author{N. Barnea}
\affiliation{\HUJI}
\author{D.W. Higinbotham}
\affiliation{\JLAB}
\author{E. Piasetzky}
\affiliation{\TAU}
\author{M. Strikman}
\affiliation{\Penn}
\author{L.B. Weinstein}
\affiliation{\ODU}
\author{O. Hen}
\email[Contact Author \ ]{(hen@mit.edu)}
\affiliation{\MIT}

\begin{abstract}
Measurements of short-range correlations in exclusive $^4$He$(e,e'pN)$ reactions are analyzed using the Generalized Contact Formalism (GCF).
We consider both instant-form and light-cone formulations with both the AV18 and local N2LO(1.0) nucleon-nucleon ($NN$) potentials.
We find that kinematic distributions, such as the reconstructed pair opening angle, recoil neutron momentum distribution, and pair center of mass motion, as well as the measured missing energy, missing mass distributions, are all well reproduced by GCF calculations.
The missing momentum dependence of the measured $^4$He$(e,e'pN)$ / $^4$He$(e,e'p)$ cross-section ratios, sensitive to nature of the $NN$ interaction at short-distacnes, are also well reproduced by GCF calculations using either interaction and formulation.
This gives credence to the GCF scale-separated factorized description of the short-distance many-body nuclear wave-function.
\end{abstract}

\maketitle

Short-range correlations (SRCs) are pairs of strongly interacting nucleons at short distance in atomic nuclei~\cite{Hen:2016kwk,Atti:2015eda}. The formation of SRCs and their exact characteristics have wide ranging implications, from the partonic structure of bound nucleons~\cite{weinstein11, Hen12, Hen:2013oha, Hen:2016kwk, Schmookler:2019nvf, Segarra:2019gbp} to the universal nature of the many-body nuclear wave-function at short-distance~\cite{Feldmeier:2011qy,ryckebusch15,Alvioli:2013qyz,CiofidegliAtti:2017xtx,Weiss:2015mba,Weiss:2016obx,Weiss:2018tbu,Cruz-Torres:2019fum}.

The seminal studies of SRCs used high-energy electron scattering to measure the hard-breakup of SRC pairs in A$(e,e'pN)$ reactions~\cite{subedi08, shneor07, korover14, hen14, duer18, Cohen:2018gzh, Duer:2018sxh}.
Two key observables in those studies are the A$(e,e'pN)$ / A$(e,e'p)$ and A$(e,e'pp)$ / A$(e,e'pn)$ cross-section ratios, which probe the isospin structure of SRC pairs.
The results of such studies established the dominance of neutron-proton ($np$) SRC pairs in the momentum range of 300 to 600 MeV/$c$~\cite{subedi08, shneor07}.  This is understood to result from the large tensor component of the $NN$ interaction in this momentum range~\cite{schiavilla07, alvioli08, sargsian05}.

At higher momentum, and thereby shorter distance, the $NN$ interaction is expected to transition from a predominantly tensor interaction to a scalar repulsive core.
This transition should lead to an increase in the fraction of proton-proton ($pp$) SRC pairs, that can be observed experimentally by an increase in the A$(e,e'pp)$ / A$(e,e'pn)$ and A$(e,e'pp)$ / A$(e,e'p)$ cross-section ratios, and a decrease in the A$(e,e'pn)$ / A$(e,e'p)$ cross-section ratio.

Ref.~\cite{korover14} searched for such a transition in $^4$He using measurements of $^4$He$(e,e'pN)$ and 
$^4$He$(e,e'p)$ reactions by small-acceptance spectrometers. The measured $^4$He$(e,e'pp)$ / $^4$He$(e,e'pn)$
ratio was generally consistent with the expected increase in $pp$-SRC pairs with increasing reconstructed
initial momentum of the knock-out nucleon. However, the extracted $^4$He$(e,e'pp)$ /  $^4$He$(e,e'p)$
ratio was consistent with no momementum-dependence.

Recently, the A$(e,e'pp)$ / A$(e,e'p)$ ratio was extracted in nuclei from $^{12}$C to $^{208}$Pb using data
from a large-acceptance spectrometer~\cite{schmidt20}. A clear increase was observed as a function of the
knock-out nucleon's initial momentum.
The data are in excellent agreement with calculations from the generalized contact formalism (GCF)~\cite{Weiss:2015mba,Weiss:2016obx,Weiss:2018tbu,Cruz-Torres:2019fum}, using both the AV18~\cite{wiringa95} and N2LO(1.0)~\cite{Gezerlis:2013ipa} potentials.

The observed increase in the A$(e,e'pp)$ / A$(e,e'p)$ ratio of Ref.~\cite{schmidt20} seems to be inconsistent with the constant ratio reported by Ref.~\cite{korover14}. However, to properly quantify the consistency of the two measurements, they need to be analyzed within the same theoretical framework that consistently accounts for the different kinematics and experimental acceptances of the two experiments.

Here we show that analyzing the two datasets with the same theoretical framework yield consistent results that support the increase of the fraction of $pp$ SRC pairs as the NN interaction changes from tensor to scalar dominance. It also contributes to the confidence of the GCF scale-separation assumption as a description of SRCs in nuclei.

In this study, we performed for the first time a GCF analysis of the $^4$He$(e,e'pN)$ and $^4$He$(e,e'p)$
measurements done by small-acceptance spectrometers.
We use the GCF in both instant-form and light-cone formulations, using both the AV18 and local N2LO(1.0) $NN$ potentials.
We find that the measured missing energy and missing mass distributions, as well as the missing momentum dependence of the $^4$He$(e,e'pN)$ / $^4$He$(e,e'p)$ cross-section ratios, are all well reproduced by GCF calculations in both formulations using either $NN$ potential.
Additional kinematic distributions, such as reconstructed pair opening angle, recoil neutron momentum distribution, and pair center of mass (c.m.) motion, are also well reproduced by the GCF.
This shows important consistency between the measurements reported here and that of Ref.~\cite{schmidt20} and gives credence to the GCF scale-separated factorized description of the short-distance many-body nuclear wave-function.


\section{Kinematics for SRC Breakup Reactions}\label{sec:Kin}
The experimental data studied here were taken using a 4.454 GeV electron beam incident on a $^4$He gas target in Hall A at Jefferson Laboratory.  Two independent small acceptance, high resolution spectrometers (HRS)~\cite{Alcorn:2004sb} were used to detect the scattered electron and knockout proton. 
Triggered by the coincidence of the two spectrometers, dedicated recoil proton and neutron detectors were used to look for their emission due to the SRC breakup reaction described below. See details in Ref.~\cite{korover14}.

This data analysis is performed within the high-resolution description of large momentum-transfer quasi-elastic (QE) 
nucleon-knockout reactions. We assume that for high initial nucleon momentum the nucleus can 
be modeled as an off-shell SRC pair with total (c.m.) momentum 
$\vec p_{cm}$, and an on-shell residual $A-2$ system. 
The electron scatters from the nucleus by exchanging a single virtual photon with 4-momentum 
($\vec q$, $\omega$) that is absorbed by a single off-shell nucleon in the SRC pair with initial 
4-momentum ($\vec{p}_1$, $E_1$).
If that nucleon does not re-scatter as it leaves the nucleus, it will emerge with momentum
 $\vec{p}\thinspace_1' = \vec{p}_1 + \vec{q}$. The measured missing momentum is defined as 
 $\vec{p}_\text{miss} = \vec{p}\thinspace_1' - \vec{q} \approx \vec{p}_1 $. 
The correlated recoil nucleon is treated as an on-shell spectator with 4-momentum 
($\vec{p}_2$, $E_2$) = ($\vec{p}_\text{CM} - \vec{p}_\text{miss}$, $\sqrt{p_2^2 + m_N^2}$),
where  $m_N$ is the nucleon mass. 
The residual $A-2$ system has momentum $-\vec{p}_\text{CM}$ and excitation energy $E^*$.

The measurements analyzed here were performed at $Q^2 =\vec q\thinspace^2 - \omega^2 \approx 2$ (GeV$/c$)$^2$
and $x_B =Q^2/2m_N\omega >$ 1.1, corresponding to anti-parallel kinematics.
While the electron spectrometer was kept fixed at these central kinematics, 
the proton spectrometer moved between three settings covering missing momentum 
ranges of [400--600], [540--720], and [660--820] MeV/$c$. See Ref.~\cite{korover14} for details. 
In these kinematics, non-QE reaction mechanisms are expected to be suppressed~\cite{Frankfurt:1996xx,frankfurt08b,Atti:2015eda,Hen:2016kwk}.
Therefore, the hard breakup of SRC pairs should provide a valid description of
the measured reactions, up to the inclusion of hard rescattering and single charge
exchange (SCX).  These effects are discussed in a later section.

\section{GCF A$(e,e'NN)$ Cross-Section}

To compare the experimental data with GCF predictions, cross-sections calculated in the GCF are used to
generate events that are processed analogously to the experimental data.  Below we present two formulations
of the GCF cross-section, followed by a description of the way they were implemented into an event generator
and compared to data.

\subsection{Instant Form Formulation}\label{subsec:NR}

The A$(e,e'N)$ nucleon-knockout cross-section for the high-$Q^2$ QE SRC breakup reaction described above
is modeled here using a factorized plane wave impulse approximation (PWIA)~\cite{DeForest:1983ahx,kelly96}:
\begin{equation}
\frac{d^6\sigma}{d\Omega_{k'}dE'_kd\Omega_{p_1'}dE'_1} = p_1'E_1'\sigma_{eN} S^N_A(p_1,E_1),
\label{eq:PWIA}
\end{equation}
where $(\vec{k}',E'_k)$ is the scattered electron four-momentum, $\sigma_{eN}$ is the off-shell electron-nucleon
cross-section~\cite{DeForest:1983ahx}, and $S^N_A(p_1,E_1)$ is the nuclear spectral function for nucleus $A$,
which defines the probability for finding a nucleon in the nucleus with momentum $p_1$ and energy $E_1$.

In the GCF, the two-body continuum region of the spectral function is given by a sum of SRC pairs with different
spin-isospin configurations~\cite{Weiss:2018tbu,CiofidegliAtti:1991mm,CiofidegliAtti:1995qe,CiofidegliAtti:2017xtx}. 
In the case of proton knockout, this amounts to:
\begin{equation}
\begin{split}
S^p(p_1,E_1)=C^1_{pn}S^1_{pn}(p_1,E_1)+&C^0_{pn}S^0_{pn}(p_1,E_1)\\
+&2C^0_{pp}S^0_{pp}(p_1,E_1),
\end{split}
\end{equation}
where $C^\alpha_{ab}$ are the nuclear contacts, which denote the probability of finding an $NN$-SRC pair with quantum numbers $\alpha$. Here $\alpha=0$ denotes a pair in a spin singlet, isospin triplet state, while $\alpha=1$ denotes a pair in a spin triplet, isospin singlet state. $S^\alpha_{ab}$ is the contribution of each channel to the total spectral function and is given by:
\begin{equation}
\begin{split}
S^\alpha_{ab}(p_1,E_1)=& \\
&\frac{1}{4\pi}\int \frac{d^3\vec{p}_2}{(2\pi)^3}
|\tilde{\phi}^\alpha_{ab}(\vec{p}_\text{rel})|^2
n^\alpha_{ab}(\vec{p}_\text{CM}) \\
& \times \delta(E_1 + E_2 + E_{A-2} - m_A),
\end{split}
\label{eq:SRCSF}
\end{equation}
where:
\begin{itemize}[leftmargin=*]
\item $\vec{p}_\text{CM} = \vec{p}\thinspace_1 + \vec{p}\thinspace_2$ and $\vec{p}_\text{rel} = \frac{\vec{p}\thinspace_1 - \vec{p}\thinspace_2}{2}$ are the c.m. and relative momentum of the pair, respectively,
\item $|\tilde{\phi}^\alpha_{ab}(\vec{p}_\text{rel})|^2$ is the universal two-body function, defining the distribution of the relative momentum of nucleons within a pair, produced by solving the two-body Schr{\"o}dinger equation for a given $NN$ potential,
\item $n^\alpha_{ab}(\vec{p}_\text{CM})=\frac{1}{(2\pi\sigma_\text{CM})^{3/2}}\exp(-\frac{\vec{p}\thinspace_\text{CM}^2}{2\sigma_\text{CM}^2})$ is the 
pair c.m. momentum distribution, taken to be a three-dimensional Gaussian with the same width ($\sigma_\text{CM}$) for all channels,
\item $E_2=\sqrt{\vec{p}\thinspace_2^2+m_N^2}$ is the energy of the spectator/partner nucleon in the pair, assumed to be on-shell,
\item $E_{A-2}=\sqrt{\vec{p}\thinspace_\text{CM}^2+(m_{A-2}+E^*)^2}$ is the energy of the residual $A-2$ system, with excitation energy $E^*$,
\item $m_A$ is the mass of the target nucleus.
\end{itemize}

Combining Eqs.~(\ref{eq:PWIA}) and~(\ref{eq:SRCSF}) we arrive at the following cross-section equation:
\begin{equation}
\label{eq:GCFXS_int}  
\begin{split}
\frac{d^6\sigma}{d\Omega_{k'}dE'_kd\Omega_{p_1'}dE'_1} &= \frac{1}{4\pi} p_1'E_1'\sigma_{eN} \\
&\int\frac{d^3\vec{p}_2}{(2\pi)^3}
\delta(W_f-W_i)\\
&\times\sum_\alpha C^\alpha_{ab}|\tilde{\phi}^\alpha_{ab}(\vec{p}_\text{rel})|^2
n^\alpha_{ab}(\vec{p}_\text{CM}),
\end{split}
\end{equation}
where $W_i = E_k + m_A$ and $W_f = E_k' + E_1' + E_2 + E_{A-2}$ are the total energies in the initial and final states respectively. Note that the $pp$-channel requires an additional factor of $2$ coming from the definition of the contact.

Eq.~(\ref{eq:GCFXS_int}) contains an integral over all possible spectator nucleon momentum with $|\vec{p}_\text{rel}|>k_\text{cut-off}$, arising from the definition of the spectral function in Eq.~(\ref{eq:SRCSF}). 
For this application of the A$(e,e'pN)$ cross-section, we need to preserve information on the spectator nucleon.  By transforming variables and integrating over the $\delta$-function, the A$(e,e'pN)$ cross-section can be expressed as:
\begin{equation}
\begin{split}
&\frac{d^8\sigma}{d\Omega_{k'}d^3\vec{p}_\text{CM}dp_\text{rel}d\Omega_\text{rel}} = \\
&\frac{\sigma_{eN}}{32 \pi^4}\frac{p_\text{rel}^2}{\left| 1 - \frac{\vec{p}\thinspace_1' \cdot \vec{k}'}{E_1'E_k'}\right|}
\sum_\alpha C^\alpha_{ab}|\tilde{\phi}^\alpha_{ab}(\vec{p}_\text{rel})|^2
n^\alpha_{ab}(\vec{p}_\text{CM}).
\label{eq:finalNR}
\end{split}
\end{equation}

\subsection{Light Cone Formulation}
Due to the high momentum of nucleons in SRC pairs, we also examine a relativistic version of the GCF based on the light cone formulation of Ref.~\cite{Frankfurt81,Frankfurt:1992ny, Artiles:2016akj}. Four-momentum vectors are expressed in light cone coordinates  in terms of plus- and minus-momentum $p^\pm \equiv p^0 \pm p^3$ as well as transverse momentum $\vec{p}_\perp \equiv (p^1,p^2)$, where the 3-component axis is aligned along the direction of the momentum transfer. It is also useful to define light-cone momentum fractions $\alpha \equiv p^-/\bar{m}$, where $\bar{m} = m_A/A$.  The average light cone fraction for a nucleon in a nucleus equals unity, and the total light cone fraction of a nucleus equals $A$. 

The light-cone formulation of the PWIA cross-section (in the two-body continuum region) is given by:
\begin{equation}
\begin{split}
&\frac{d^9\sigma}{dE_k' d\Omega_{k'}\frac{d\alpha_1}{\alpha_1}d^2\vec{p}_{1,\perp}\frac{d\alpha_{2}}{\alpha_{2}}d^2\vec{p}_{{2},\perp}}=\\
&\sigma_{eN}\delta(W_f-W_i)\frac{\rho(\alpha_1,\vec{p}_{1,\perp},\alpha_{2},\vec{p}_{{2},\perp})}{\alpha_1}
\label{eq:PWIALC}
\end{split}
\end{equation}
where the two-nucleon density matrix $\rho(\alpha_1,\vec{p}_{1,\perp},\alpha_2,\vec{p}_{2,\perp})$ 
can be written in a factorized form of:
\begin{equation}
\begin{split}
&\rho(\alpha_1,\vec{p}_{1,\perp},\alpha_{2},\vec{p}_{{2},\perp})=\\
&\frac{\alpha_{2}}{\alpha_\text{CM}}\rho_\text{SRC}(\alpha_\text{rel},\vec{p}_{\text{rel},\perp})\rho_\text{CM}(\alpha_\text{CM},\vec{p}_{\text{CM},\perp}).
\end{split}
\end{equation}

Here we define the relative and c.m. momentum fractions:
\begin{equation}
\begin{split}
\alpha_\text{CM} &= \alpha_1 + \alpha_2, \\
\vec{p}_{\text{CM},\perp} &= \vec{p}_{1,\perp} + \vec{p}_{2,\perp}, \\
\alpha_\text{rel} &= \frac{2\alpha_2}{\alpha_\text{CM}}, \\
\vec{p}_{\text{rel},\perp} &= \vec{p}_{2,\perp} - \frac{\alpha_2}{\alpha_\text{CM}}\vec{p}_{\text{CM},\perp} \\
 &= \frac{\alpha_1 \vec{p}_{2,\perp} - \alpha_2 \vec{p}_{1,\perp}}{\alpha_\text{CM}}.
\label{eq:relperp}
\end{split}
\end{equation}

We note that $\vec{p}_{\text{rel},\perp}$ is not simply the perpendicular component
of $\vec{p}_{rel}$, but is adjusted for boost effects~\cite{piasetzky06}.

The density matrix for the pair relative motion is given by~\cite{piasetzky06}:
\begin{equation}
\rho_\text{SRC}(\alpha_\text{rel},\vec{p}_{\text{rel},\perp})=\sum_\alpha C^\alpha_{ab}\frac{\sqrt{m_N^2+k^2}}{2-\alpha_\text{rel}}
\frac{|\tilde{\phi}^\alpha_{ab}(k)|^2}{(2\pi)^3},
\end{equation}
where
\begin{equation}
k^2 \equiv \frac{m_N^2 + \vec{p}\thinspace_{\text{rel},\perp}^2}{\alpha_\text{rel}(2-\alpha_\text{rel})} - m_N^2.
\end{equation}
The density matrix for the pair c.m. motion is modeled by a three-dimensional Gaussian~\cite{Cohen:2018gzh}:
\begin{equation}
\begin{split}
&\rho_\text{CM}(\alpha_\text{CM},\vec{p}_\text{CM}) = \\
&\frac{\bar{m}\alpha_\text{CM}}{(2\pi\sigma_\text{CM})^{3/2}}
\exp\left\{-\frac{\bar{m}^2(2-\alpha_\text{CM})^2+\vec{p}\thinspace_{\text{CM},\perp}^2}{2\sigma_\text{CM}^2}\right\}.
\end{split}
\end{equation}

By transforming variables and integrating over the $\delta$-function, Eq.~\ref{eq:PWIALC} can be expressed similarly to Eq.~\ref{eq:finalNR}:
\begin{equation}
\begin{split}
&\frac{d^8\sigma}{d\Omega_{k'}d^3\vec{p}_\text{CM}dp_\text{rel}d\Omega_{rel}} = \\
&\frac{\sigma_{eN}}{4 \pi \alpha_1}\frac{p_\text{rel}^2}{\left|1 - \frac{\vec{p}\thinspace_1' \cdot \vec{k}'}{E_1'E_k'}\right|}
\frac{1}{E_2}\rho_\text{SRC}(\alpha_\text{rel},\vec{p}_{\text{rel},\perp})\\
&\times\frac{\alpha_{A-2}}{\alpha_\text{CM}E_{A-2}}\rho_\text{CM}(\alpha_\text{CM},\vec{p}_{\text{CM},\perp})\label{eq:finalLC}
\end{split}
\end{equation}

\begin{figure}[tp]
\includegraphics[width=\columnwidth]{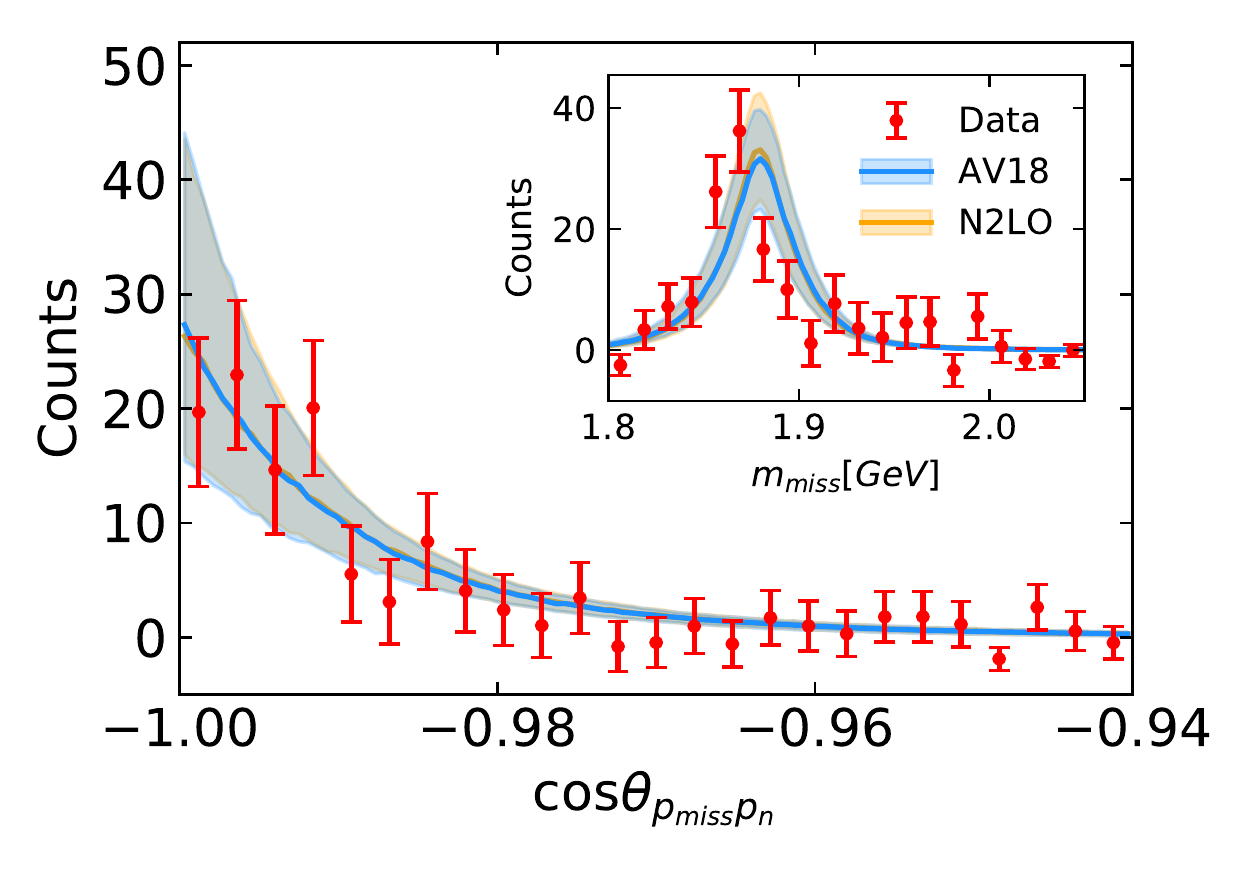}
\caption{Measured and GCF-calculated event yield distribution of the cosine of the opening angle between $\vec{p}_{recoil}$ and $\vec{p}_{miss}$ for $^4$He$(e,e'pn)$ events. Insert: same for the missing mass distribution. See Sec.~\ref{sec:Results} for details.}
\label{Fig:kin}
\end{figure}

\section{\label{sec:tech}GCF Event Generator Implementation}


To compare with experimental data, the cross-section expressions of Eq.~\ref{eq:finalNR}
and~\ref{eq:finalLC} are used to produce a weighted Monte Carlo event generator. 
We further model radiative and reaction mechanism effects, and then propagate 
the resulting pseudo-events through a model of the experiment. The procedure
is described in the following subsections.

\subsection{\label{subsubsec:gen}Event Generation and Kinematics}

As we have specified our cross-sections to be differential in $\Omega_{k'}$, ${\bf p}_{CM}$, $p_{rel}$,
and $\Omega_{rel}$, we randomly sample our generated kinematics in these variables according to the 
probability distribution:
\begin{multline}\label{eq:dist}
P(\Omega_{k'},\vec{p}_\text{CM},p_\text{rel},\Omega_\text{rel})\\
=\frac{1}{\Delta \Omega_{k'}}\times n(\vec{p}_{CM})\times\frac{1}{4\pi}\times\frac{1}{\Delta p_\text{rel}},
\end{multline}
i.e., $\Omega_{k'}$, $p_\text{rel}$, and $\Omega_\text{rel}$ are sampled from independent uniform distributions,
restricted to regions allowed by the spectrometer acceptance, and $\vec{p}_\text{CM}$ is sampled from a Gaussian
distribution of width $\sigma_\text{CM}$.  After selecting these variable, $E_{k'}$ can be determined from energy conservation 
(i.e. $m_A + \omega = E_1' + E_2 + E_{A-2}$).
The recoil nucleon is selected randomly to be either a proton or a neutron with the corresponding form-factors
used for the off-shell electron-nucleon cross-section calculation.

\subsection{\label{subsubsec:weight}Event Weighting}

Each pseudo-event is assigned a weight, given by
\begin{equation}
w=\frac{d\sigma (\Omega_{k'},\vec{p}_\text{CM}, p_\text{rel},\Omega_\text{rel}) }{P(\Omega_{k'},\vec{p}_\text{CM}, p_\text{rel},\Omega_\text{rel})}
\end{equation}
where $d\sigma$ is the differential cross section for the event's kinematics, and $P$ 
probability for sampling the event's kinematics. Using Eqs.~\ref{eq:finalNR} and~\ref{eq:dist},
the instant-form PWIA weight is
\begin{equation}
w_{IF} = \frac{\sigma_{eN}}{8 \pi^3}\Delta \Omega_{k'}\frac{p\thinspace_\text{rel}^2 \Delta p_\text{rel}}
{\left| 1 - \frac{\vec{p}\thinspace_1' \cdot \vec{k}'}{E_1'E_k'}\right|}
\sum_\alpha C^\alpha_{ab}|\tilde{\phi}^\alpha_{ab}(\vec{p}_\text{rel})|^2.
\label{eq:weightNR}
\end{equation}
The light cone version (using Eqs.~\ref{eq:finalLC} and~\ref{eq:dist}) is
\begin{equation}
\begin{split}
w_{LC} = &\frac{\sigma_{eN}}{\alpha_1}\Delta \Omega_{k'}\frac{p_\text{rel}^2\Delta p_\text{rel}}
{\left|1 - \frac{\vec{p}\thinspace_1' \cdot \vec{k}'}{E_1' E_k'}\right|}
\frac{1}{E_2}\rho_\text{SRC}(\alpha_\text{rel},\vec{p}_{\text{rel},\perp})  \\
&\times\frac{\bar{m}\alpha_{A-2}}{E_{A-2}} \exp\left\{\frac{\vec{p}_{\text{CM},||}^2-\bar{m}^2(2-\alpha_\text{CM})^2}{2\sigma_\text{CM}^2}\right\}.
\label{eq:weightLC}
\end{split}
\end{equation}

\begin{figure*}[htp]
\includegraphics[width=0.68\columnwidth]{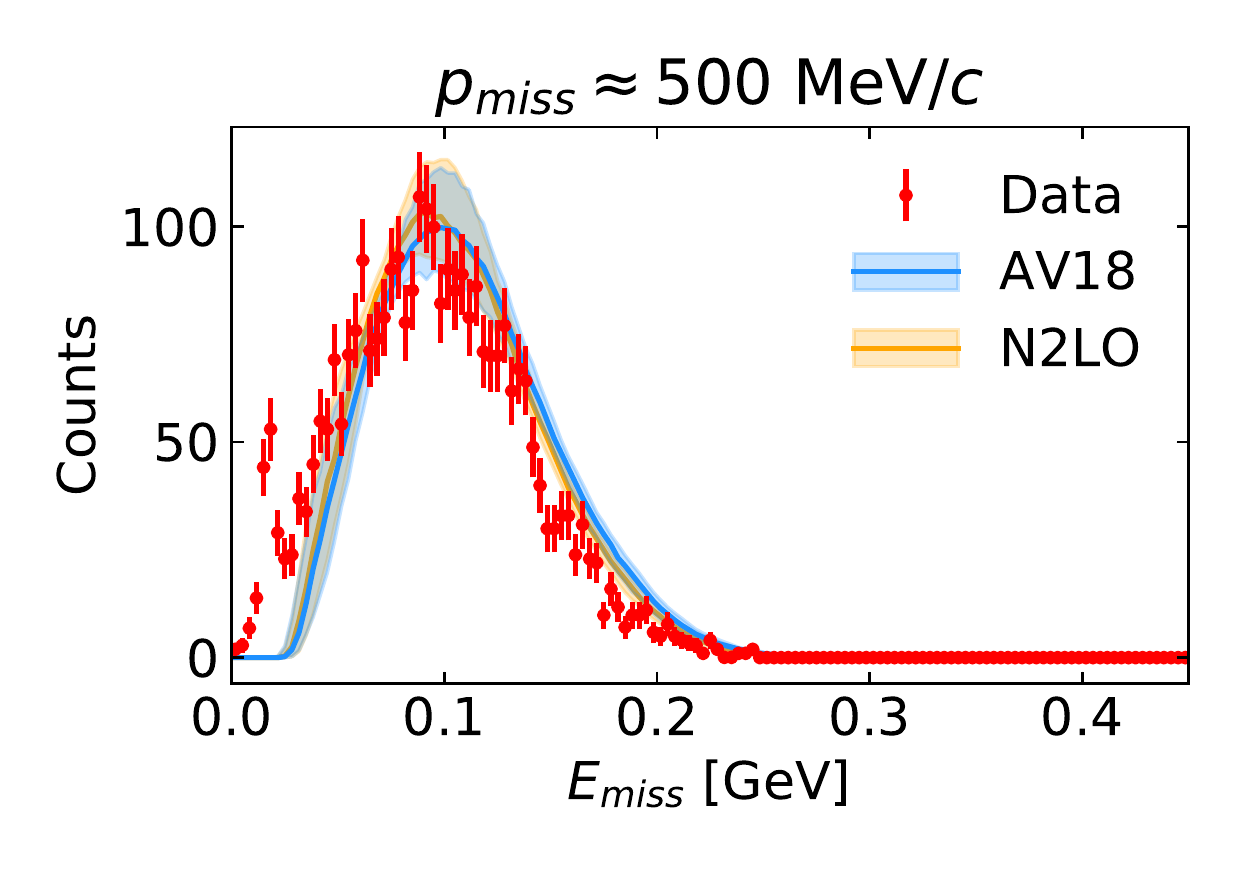}
\includegraphics[width=0.68\columnwidth]{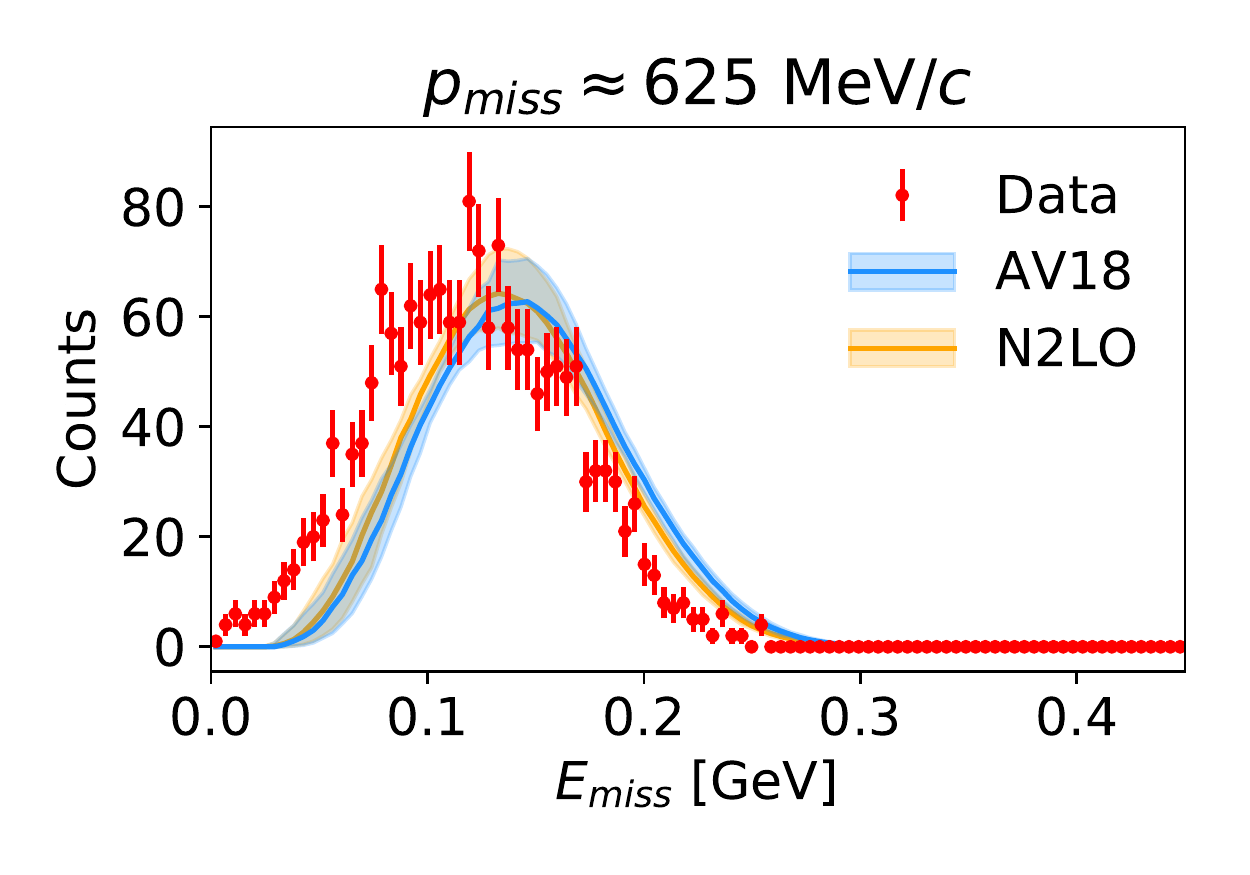}
\includegraphics[width=0.68\columnwidth]{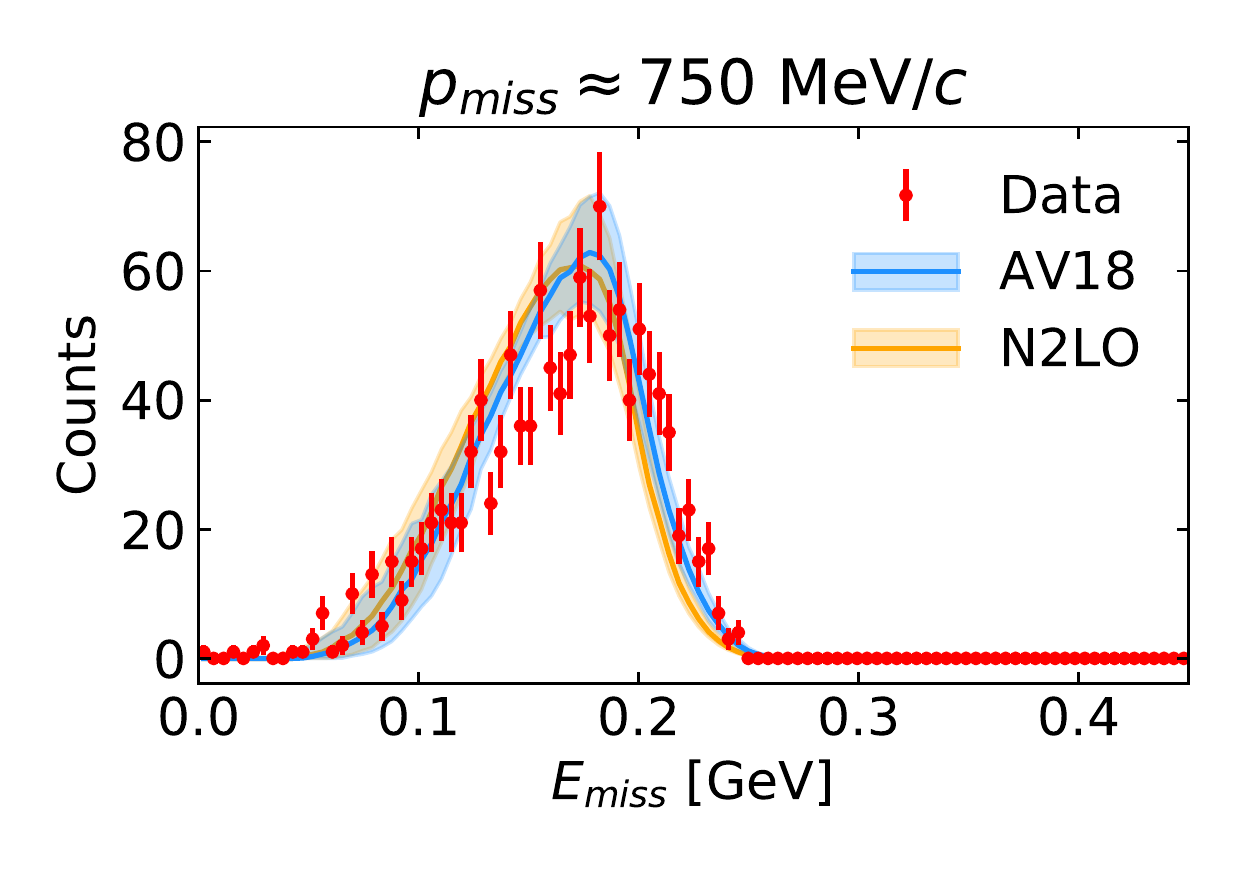}
\caption{Missing energy dependence of measured $^4$He$(e,e'p)$ event yields~\cite{korover14} for three kinematical settings compared with GCF calculations. Kinematical settings have increasing central missing momentum from left to right. See Sec.~\ref{sec:Results} for details.}
\label{Fig:Em}
\end{figure*}

\subsection{\label{subsubsec:rad}Radiative Effects}

Comparison with measured electron scattering data requires accounting for radiative effects
beyond the Born approximation. We use a Monte Carlo approach similar to those proposed in
Ref.~\cite{Ent:2001hm}, employing the peaking approximation---energy radiated by bremsstrahlung
is only emitted in the incoming and outgoing electron directions---as well using exponentiation
to describe the multi-photon radiated energy distribution. First, the energy radiated by the 
incoming electron and the energy radiated by the outgoing electron are randomly sampled
according to the probability distribution:
\begin{equation}
P(E_\text{rad.}) = \frac{\lambda}{E_{k^{(')}}}\left( \frac{E_\text{rad.}}{E_{k^{(')}}} \right)^{\lambda - 1},
\end{equation}
\begin{equation}
\lambda = \frac{\alpha}{\pi}\left[\log \left( \frac{4E_{k^{(')}}^2}{m_e^2} \right) - 1 \right],
\end{equation}
where $E_\text{rad.}$ is the total energy radiated by an electron leg in the Feynman diagram, 
$E_{k^{(')}}$ is the energy carried by the electron leg prior to radiation, $m_e$ is the electron mass, and $\alpha$
is the fine-structure constant. The GCF cross-section is calculated using the modified electron
kinematics, i.e., after initial state radiation but before final state radiation. The event
weights are multiplied by a further radiative correction factor given by
\begin{equation}
w_\text{rad.}=\left(1 - \delta_\text{hard} \right) \times \left( \frac{E_k}{\sqrt{E_k E_{k'}}} \right)^{\lambda_{i}} 
\times \left( \frac{E_k + E_\text{rad.}^f}{\sqrt{E_k E_{k'}}}\right)^{\lambda_{f}},
\end{equation}
with
\begin{equation}
\delta_\text{hard} = \frac{2 \alpha}{\pi} \left[ \frac{-13}{12}\log\left(\frac{Q^2}{m_e^2}\right) + \frac{8}{3}\right].
\end{equation}
This approach to radiative corrections is equivalent to the ``pure peaking approximation'' approach
of Ref.~\cite{Ent:2001hm}, but further neglecting non-peaked bremsstrahlung strength and bremsstrahlung
from any nucleon.

\subsection{Reaction Mechanism Effects}
\label{subsubsec:FSI}
Following Refs.~\cite{hen14,Duer:2018sxh,schmidt20}, we account for the main reaction effects relevant for the kinematics of the data being analyzed here. Due to the anti-parallel nature of the measured reaction, these include flux reduction due to hard rescattering (Transparency) and isospin changes in the final state due to $(n,p)$ and $(p,n)$ SCX reactions. 

We account for these effects by constructing an approximate `experimental equivalent' cross-section expressions from the GCF PWIA calculated cross-sections, e.g.:
\begin{equation} 
\begin{split}
\sigma^{Exp}_{A(e,e'pN)} =& \sigma^{GCF}_{A(e,e'pN)} \cdot P_{A}^{pN}\cdot T_{A} + \\
& \sigma^{GCF}_{A(e,e'nN)}\cdot P_{A}^{[n]N}\cdot T_{A} + \\
& \sigma^{GCF}_{A(e,e'pN')}\cdot P_{A}^{p[N']}\cdot T_{A},
\end{split}
\label{eq:scx_gcf}
\end{equation}
Where $T_{A}$ and $P_A$ are respectively Transparency and SCX probabilities, taken from reaction calculations~\cite{Colle:2015lyl}, which agree well with experimental data~\cite{hen12a,Duer:2018sjb,colle15}. The use of `$[N]$' in the SCX supscript marks the nucleon in the pair that undergoes SCX into a different isospin state. We assume that the transparency of nucleons following SCX is the same as for nucleons that did not undergo SCX. We further note that the single nucleon transparency is calculated to be only slightly larger than that of a pair of nucleons. See Ref.~\cite{igor:Thesis, Duer:2018sxh} for details. 

All comparisons to data in this work are made using the `experimental equivalent' cross-sections defined here.

\begin{figure*}[htp]
\includegraphics[width=0.68\columnwidth]{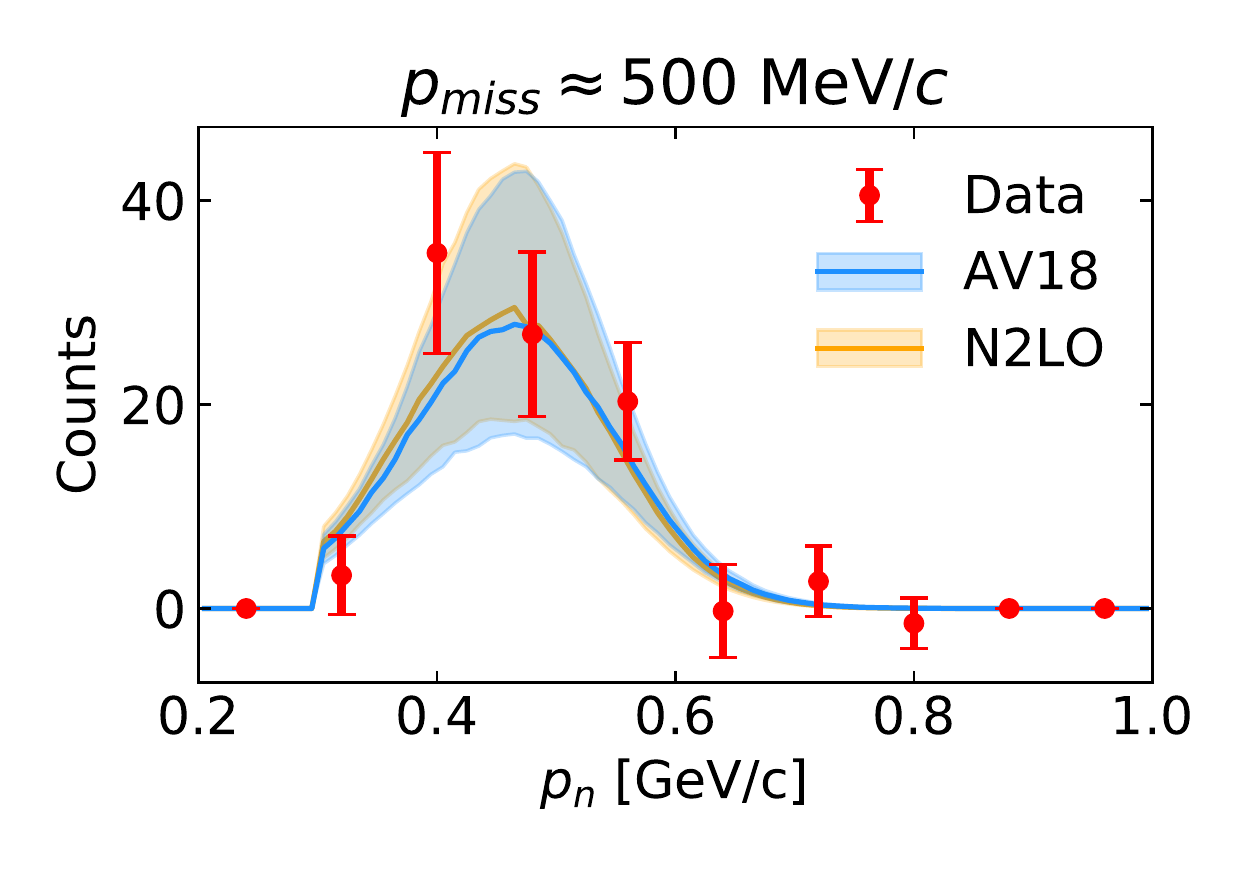}
\includegraphics[width=0.68\columnwidth]{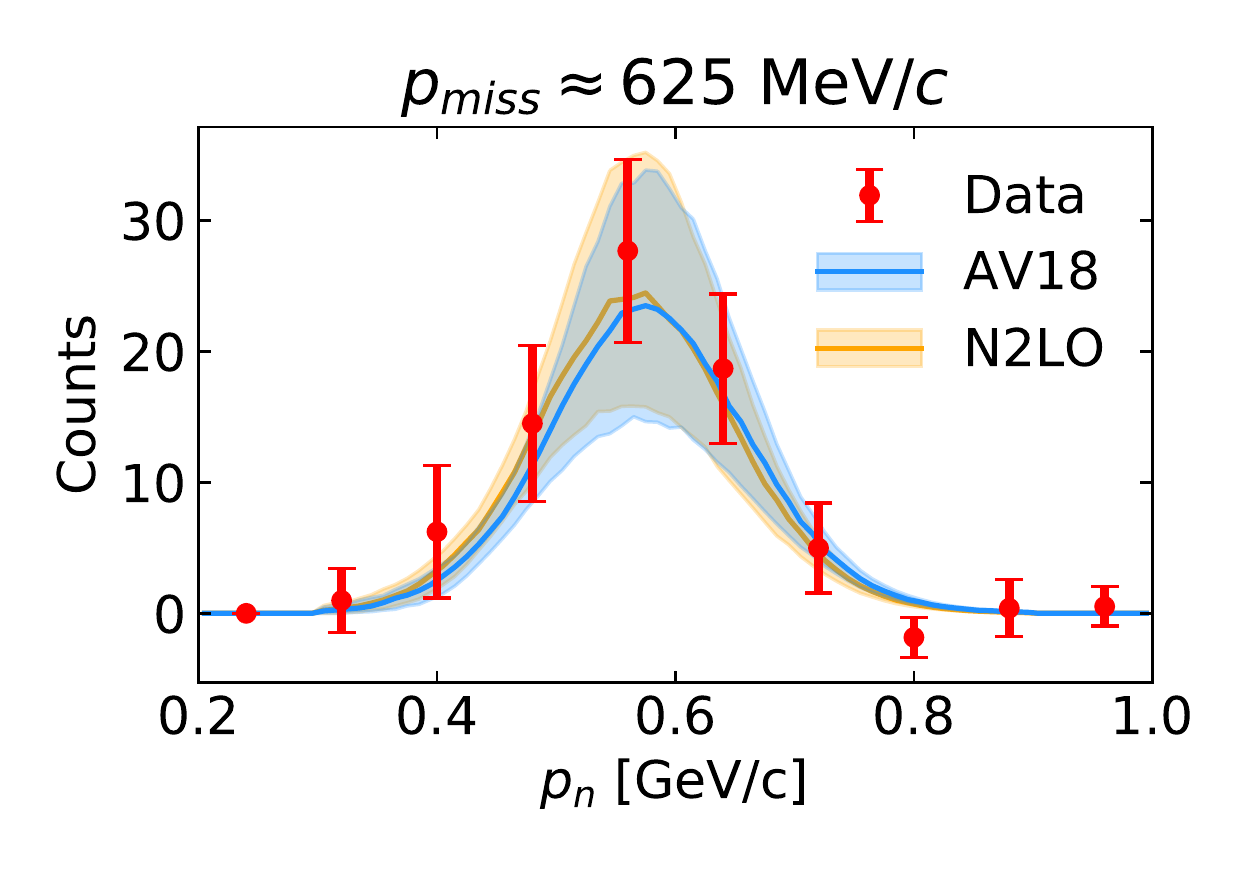}
\includegraphics[width=0.68\columnwidth]{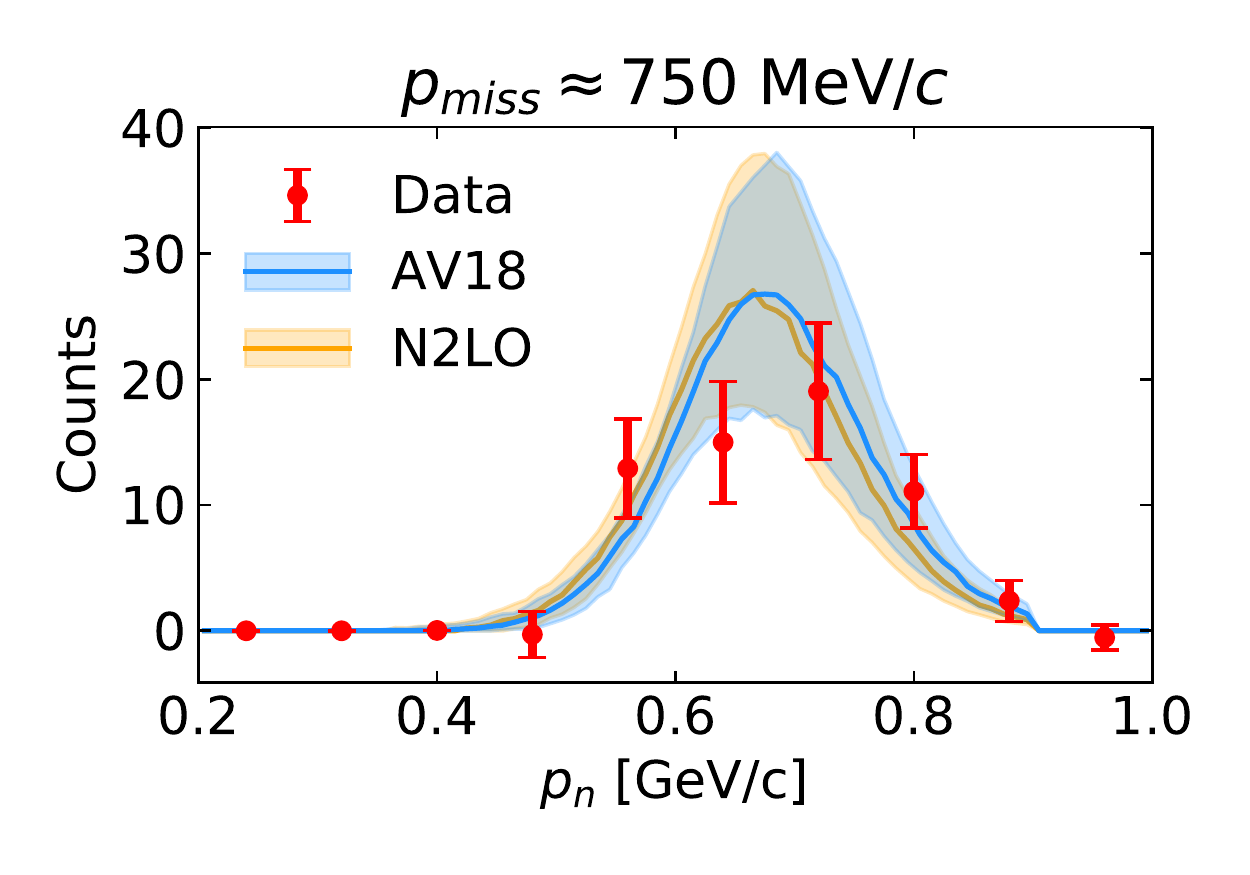}
\caption{Recoil neutron momentum distribution for measured $^4$He$(e,e'pn)$ event yields~\cite{korover14} for three kinematical settings compared with GCF calculations. Kinematical settings have increasing central missing momentum from left to right. See Sec.~\ref{sec:Results} for details.}
\label{Fig:prec}
\end{figure*}

\subsection{Model Systematic Uncertainties}

The cross-section Eqs.~\ref{eq:finalNR},~\ref{eq:finalLC} and~\ref{eq:scx_gcf} require several input parameters. 
While their values have been determined by previous works, their uncertainty leads to an uncertainty in the 
calculated cross-section. We estimate this uncertainty by performing the calculation many times, while simultaneously
varying all of the input parameters according to a prior probability distribution. For the results shown in this work,
we indicate the median value of the calculations as well as a band which contains 68\% of the sample parameter
combinations.

The following parameters were varied according to a Gaussian distribution unless otherwise indicated:
\begin{itemize}[leftmargin=*]
\item $\sigma_{CM}$, the width of the SRC pair c.m. momentum distribution, which was assumed to 
equal $100\pm 20$ MeV$/c$, as extracted from the original analysis~\cite{korover14},
\item $C^\alpha_{ab}$, the nuclear contacts, which were taken from momentum-space VMC calculations in Ref.~\cite{Cruz-Torres:2019fum},
\item $P^{SCX}_A = 1.5\pm1.5\%$, the SCX probability, which was taken from the original 
analysis~\cite{korover14,igor:Thesis}, with negative values excluded,
\item $T_A = 0.7$, the nuclear transparency, which was taken from the original analysis~\cite{korover14,igor:Thesis} 
with an assumed $\pm 20\%$ uncertainty,
\item $k_\text{cut-off}$, the momentum cut-off in the universal two-body function above which SRCs begin to dominate,
which was varied from a uniform distribution between 200--300~MeV$/c$,
\item $E^*$, the excitation energy of the residual $A-2$ nucleus, which was varied uniformly between 0--10~MeV.
\end{itemize}

\begin{figure*}[t]
\includegraphics[width=0.75\columnwidth]{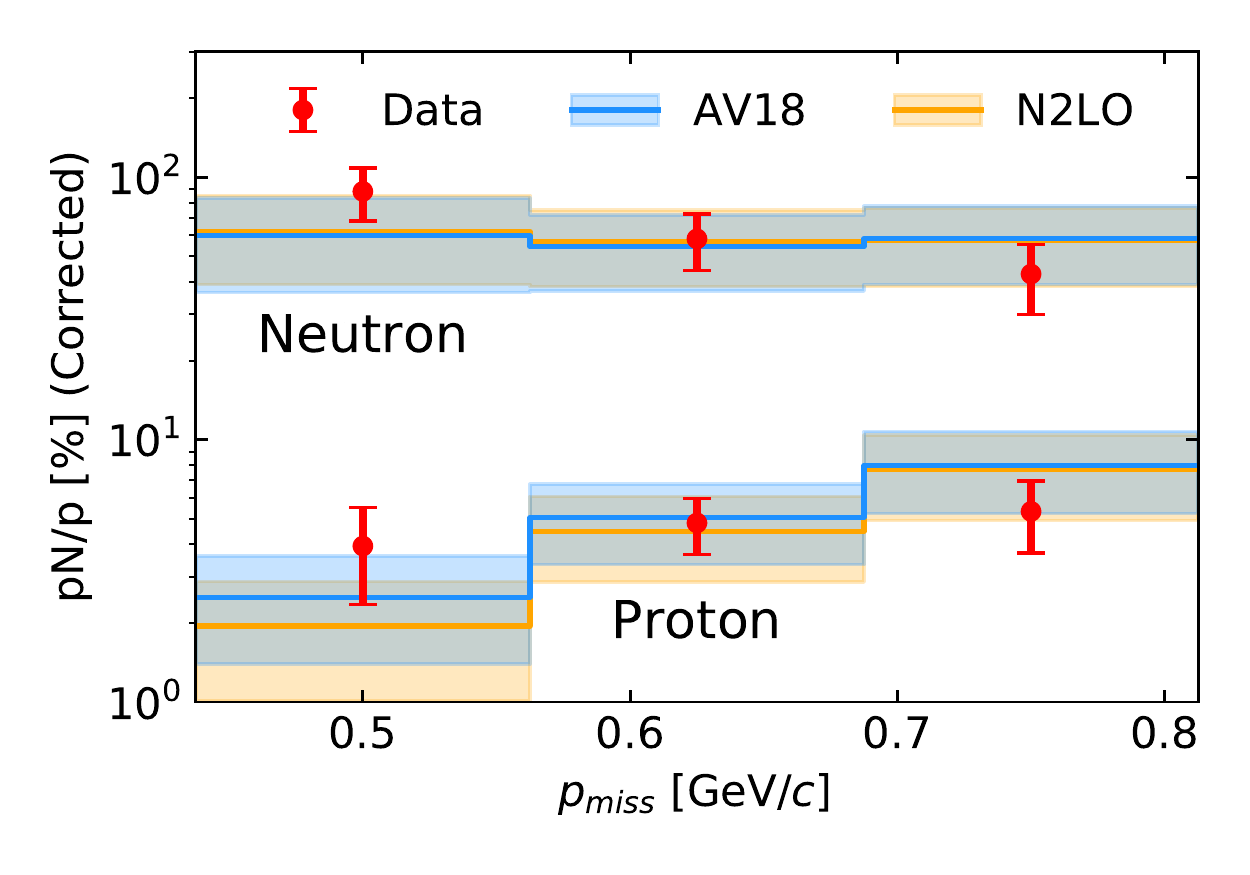}
\includegraphics[width=0.75\columnwidth]{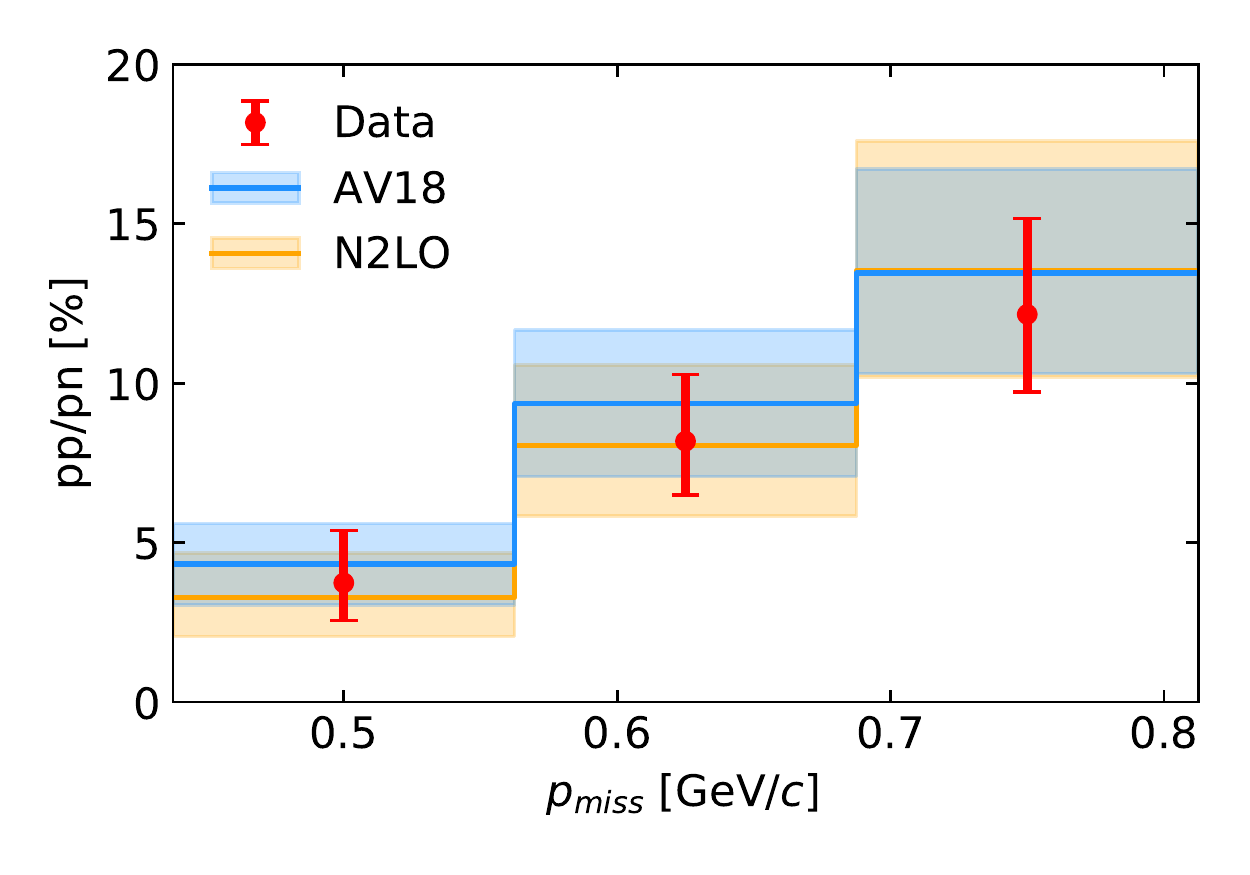}
\caption{Left: Cross section ratios $^4$He$(e,e'pN)$ / $^4$He$(e,e'p)$ for Ref.~\cite{korover14} and GCF calculations. Ratios were taken for 3 kinematical settings centered around 3 bins in missing momentum. Includes ratios with recoil neutron (top) and recoil proton (bottom).
Right: Event yield super-ratio $^4$He$(e,e'pp)$ / $^4$He$(e,e'pn)$ and GCF calculations across kinematical settings.}
\label{Fig:pN_p}
\end{figure*}

The systematic uncertainty bands presented in Figs.~\ref{Fig:kin}--\ref{Fig:pN_p} account for correlated effects through
simultaneous variation all model parameters. The impact of each individual model parameter can be found
in online supplementary materials tables I--IV, though these estimates necessarily neglect correlated effects.


\subsection{Event Selection and Comparison with Data}
Pseudo-events from the event generator were analyzed in an identical fashion to the events measured in 
the experimennt. We applied a model for the spectrometer acceptances to reject any pseudo-events that would
not have been triggered during the experiment. We then applied the same event selection criteria as in the
experimental analysis:
\begin{itemize}[leftmargin=*]
\item Scattered electron and leading proton were in the fiducial region of the HRSs: In-plane angle $\pm 30$ mrad,
out-of-plane angle $\pm 60$ mrad, and momentum acceptance $\pm 4.5$ \%,
\item Recoil nucleon was in the fiducial region of BigBite/HAND: In-plane angle $\pm 14^\circ$,
out-of-plane angle $\pm 4^\circ$, and momentum within $300-900$ MeV$/c$,
\item A linear cut on energy transfer $\omega$ and the $y$-scaling variable, $\omega<A y+B$, with $A = -1.32, -1.28, -1.25$ and $B = 0.90, 0.88, 0.86$ in the three kinematical settings, respectively,
\item Cut on the missing energy, $E_{miss} = m_N - m_A + \sqrt{(\omega+m_A - E_{lead})^2-\vec{p}_{miss}^2} >30$ MeV,
\item Cut on the missing mass, $m_{miss} = \sqrt{(\omega+2m_N - E_{lead})^2 - \vec{p}_{miss}^2} <1$ GeV$/c^2$, for events with a detected recoil nucleon, only in the $p_{miss}\approx$ 750 MeV$/c$ kinematic setting,
as detailed in Ref.~\cite{igor:Thesis},
\end{itemize}
As detector inefficiencies were corrected for in the original analysis, we did not apply any efficiency
corrections to the calculation.

Kinematical distributions shown in Ref.~\cite{korover14,igor:Thesis} are reported as `event yield'
distributions, not as cross-sections. Our treatment of the event generator pseudo-data allows us 
to make comparisons on equal footing, up to the limit of an overall normalization factor for
each kinematical setting. We have chosen to normalize the calculation to the yield of measured
$^4$He$(e,e'p)$ events for each kinematical setting. This choice automatically determines the 
normalization of calculated $^4$He$(e,e'pN)$ yields. For the lowest $\vec{p}_{miss}$ kinematics,
we excluded low missing-energy two-body breakup from this normalization procedure, since this is
outside the purview of GCF. We note that the normalization factors cancel in the $^4$He$(e,e'pN) /^4$He$(e,e'p)$
and $^4$He$(e,e'pp)/^4$He$(e,e'pn)$ ratios. The normalization constants for AV18 and N2LO calculations
differ by factors of 1.06, 0.78, and 0.52 for the $p_{miss}\approx$ 500, 625, and 750 MeV$/c$ settings, respectively.
This means that if Ref.~\cite{korover14,igor:Thesis} were to report absolute cross-sections one of the models,
most likely N2LO, would not manage to describe its decrease with missing-momentum.

\section{\label{sec:Results}Results}

As instant form and light cone results are very similar, here we only show results for the former while the latter are shown in the online supplementary materials. Future measurements, beyond the scope of the data analyzed here, can have an enhanced sensitivity to relativistic effects by exploring a wide-range of kinematical correlations that can highlight differences between the two approaches.

Fig.~\ref{Fig:kin} shows the measured and GCF-calculated event yield distribution of the cosine of the opening angle of the pair,
i.e., the angle between $\vec{p}_{recoil}$ and $\vec{p}_{miss}$, for $^4$He$(e,e'pn)$ events
($p_{miss}\approx$ 625 and 750 MeV$/c$ kinematic settings combined).
The insert shows the missing mass distribution for the same events.
The missing mass distribution for $^4$He$(e,e'pp)$ events is shown
in online supplementary materials Fig. 5.

Fig.~\ref{Fig:Em} and~\ref{Fig:prec} respectively show the measured event yield missing energy distribution for $^4$He$(e,e'p)$ events and recoil neutron momentum distribution for $^4$He$(e,e'pn)$ events for each measured kinematical setting.
As can be seen, all measured event yield distributions are overall well described by the GCF calculations, within uncertainties.
As expected, for the lowest $\vec{p}_{miss}$ kinematics the calculated missing energy distribution do not show a two-body breakup peak as the data. In addition the missing-energy distribution for the mid  $\vec{p}_{miss}$ kinematics is slightly shifted as compared with the data.

Fig.~\ref{Fig:pN_p} shows the measured $^4$He$(e,e'pp)$ / $^4$He$(e,e'pn)$ (right) and $^4$He$(e,e'pN)$ / $^4$He$(e,e'p)$ (left) ratios as a function of missing momentum compared with GCF calculations.
Unlike the measured event yields, the $^4$He$(e,e'pN)$ / $^4$He$(e,e'p)$ ratios were corrected for the recoil nucleon acceptance.
The original correction was done using a simple phenomenological, data-driven, model. Using the GCF we independently calculated this correction factor to find that it is in excellent agreement with that used in the original analysis  (see online supplementary materials Fig. 6).
The data are consistent with GCF predictions within uncertainties for both $^4$He$(e,e'pp)$ and  $^4$He$(e,e'pn)$ reaction, and especially for their ratio.

The agreement of the GCF calculation with the seemingly constant experimental measurement of $^4$He$(e,e'pp)$ / $^4$He$(e,e'p)$ is encouraging. It shows that there is no contradiction between the spectrometer data analyzed here and the large-acceptance detector measurements of Ref.~\cite{schmidt20}.  Rather, it highlights the need for proper theoretical framework to properly account for phase-space and acceptance effects in the different measurements before relating the measured observables to ground state properties of nuclei.

The improved agreement of the $^4$He$(e,e'pp)$ / $^4$He$(e,e'pn)$ ratio data further supports previous claims that ratios of two-nucleon knockout reactions are good observables.  Such ratios not only benefit from the cancellation of many experimental uncertainties, but also from the cancellation of amplitude-level FSI.  The latter have previously been found to have significant effects in QE scattering in light nuclei~\cite{Cruz-Torres:2019fum}.

We further observe that both the AV18 and N2LO $NN$ interaction models are capable of explaining the data up to very high values of missing momentum, giving credence to their use in calculations of high-density nuclear systems.

Last, the GCF calculation additionally allows exploring the underlying pair relative momentum distribution probed in each kinematical setting. These distributions are shown in online supplementary materials Fig. 7 and 8. They are similar for the AV18 and N2LO $NN$ interaction models and for light-front and instant form GCF formulations. In all cases the pair relative momentum distribution is smaller than the probed $|\vec{p}_{miss}|$, due to the pair c.m. motion. At the lowest $|\vec{p}_{miss}|$ value the probed relative momentum distribution for the  $^4$He$(e,e'p)$ reaction is slightly shifted to lower values as compared with that of the  $^4$He$(e,e'pN)$ reactions.

\section{\label{sec:Summary}Summary}
We performed a re-analysis of SRC studies using the $^4$He$(e,e'p)$ and $^4$He$(e,e'pN)$ reactions.
The data are taken at high-$Q^2$, $x_B > 1$, high-$\vec{p}_{miss}$ kinematics that are dominated by the hard breakup of nucleons in SRC pairs.
GCF calculations of the measured reactions were done using a dedicated event generator with both instant form and light-cone formulations, while accounting for the measurement experimental setup, event selection criteria, and Transparency and SCX reaction effects.

Overall good agreement is observed between the data and GCF, especially for $^4$He$(e,e'pp)$ / $^4$He$(e,e'pn)$ and $^4$He$(e,e'pp)$ / $^4$He$(e,e'p)$ ratios. These observations give further credence for the GCF modeling of the correlated part of the nuclear ground state and the validity of the $NN$ interaction models examined here in describing two-body interactions at high-momentum and short-distances.
Future studies of three-nucleon correlations will allow extending this study of $NN$ interactions to short-distance $NNN$ interactions that are of high-interest for complete and accurate modeling of the nuclear symmetry energy at high-densities and the cooling rate of neutron stars~\cite{Gandolfi:2011xu,Li:2018lpy,hen15,Vidana11,frankfurt08b}.

\begin{acknowledgements}
This work was supported by the U.S. Department of Energy, Office of Science, Office of Nuclear Physics under Award Numbers DE-FG02-94ER40818, DE-SC0020240, DE-FG02-96ER-40960, DE-FG02-93ER40771, and DE- AC05-06OR23177 under which Jefferson Science Associates operates the Thomas Jefferson National Accelerator Facility, the Israeli Science Foundation (Israel) under Grants Nos. 136/12 and 1334/16, the Pazy foundation, and the Clore Foundation.
\end{acknowledgements}

\bibliography{../../../../references.bib}

\end{document}